\documentclass[conference,letterpaper,10pt]{IEEEtran}

\title{On the Observability of Gaussian Models\\
 using Discrete Density Approximations}
\author{
	\IEEEauthorblockN{
		\textbf{Ariane Hanebeck and}
		\textbf{Claudia Czado}}
  \IEEEauthorblockA{
		Applied Mathematical Statistics and Munich Data Science Institute (MDSI)\\
		TUM Department of Mathematics\\
		Technical University of Munich (TUM), Germany\\
		ariane.hanebeck@tum.de, czado@tum.de}}

\usepackage{amsmath} 
\usepackage{amssymb}
\usepackage{amsthm}
\usepackage[T1]{fontenc} 
\usepackage[utf8]{inputenc}
\usepackage{rotating}
\usepackage{todonotes}
\usepackage{accents}
\usepackage{graphicx,scalerel}
\usepackage{xcolor,colortbl}

\usepackage{balance}

\usepackage{lscape}
\usepackage{pdflscape}
\usepackage{longtable}
\usepackage{tabularx}
\newcolumntype{C}[1]{>{\centering\arraybackslash}p{#1}} 

\usepackage{tablefootnote}

\makeatletter
\newcommand\footnoteref[1]{\protected@xdef\@thefnmark{\ref{#1}}\@footnotemark}
\makeatother

\usepackage{makecell}

\usepackage{listings}
\usepackage{graphicx}
\usepackage{array}
\usepackage{microtype}
\usepackage{hyperref}
\usepackage{cleveref}
\usepackage{nameref}
\usepackage{footmisc}
\usepackage{caption}

\makeatletter
\Hy@AtBeginDocument{%
  \def\@pdfborder{0 -1 1}
}
\makeatother

\makeatletter
\g@addto@macro\bfseries{\boldmath}
\makeatother

\usepackage[ruled, noline]{algorithm2e}
\SetKwInput{KwIn}{In}
\SetKwInput{KwOut}{Out}
\SetAlCapSkip{1em}
\SetAlFnt{\small}
\SetAlCapHSkip{0pt}



\def\equationautorefname#1#2\null{Eq.#1(#2\null)}

\def\sectionautorefname#1\null{Section#1\null}

\numberwithin{equation}{section} 

\newcommand{\R}{\mathbb{R}} 
\newcommand{\E}{\mathbb{E}} 
\newcommand{\bmu}{\boldsymbol\mu} 

\DeclareMathOperator{\Var}{Var}
\DeclareMathOperator*{\argmax}{argmax}
\DeclareMathOperator{\LVar}{LVar}

\newlength\EqLen

\def\ScaleInner#1{%
\settowidth{\EqLen}{#1}
\ifdim\EqLen < \columnwidth%
  \begin{equation*}%
    \begin{minipage}{\EqLen}#1\end{minipage}%
  \end{equation*}%
\else%
  \begin{equation*}%
    \resizebox{0.99\columnwidth}{!}{\begin{minipage}{\EqLen}#1\end{minipage}}%
  \end{equation*}%
\fi%
}%

\def\Scale#1
  {
  \ScaleInner{$ #1 $} 
  }

\newcommand{\SingleFigureHereSpalte}[4]{%
	\begin{figure}[t]
		\begin{center}
      	\includegraphics[width=#1]{Figures/#2.pdf}
		\vspace*{-5mm}
		\end{center}
		\caption{#3}
		\label{#4}
	\end{figure}
}%

\newcommand{\DoubleFigureHere}[3][8cm]{%
	\begin{figure*}[hh]
		\hspace*{\fill}
		\includegraphics[width=#1]{Figures/#2_1.pdf}
		\hspace*{\fill}
		\includegraphics[width=#1]{Figures/#2_2.pdf}
		\hspace*{\fill}
		\caption{#3}
		\label{#2}
	\end{figure*}
}%

\newcommand{\ThreeFigureDifferentHere}[5][5.5cm]{%
\begin{figure*}[t]
	\hspace*{\fill}
	\includegraphics[width=#1]{Figures/#2.pdf}
	\hspace*{0cm}
	\includegraphics[width=#1]{Figures/#3.pdf}
	\hspace*{0cm}
	\includegraphics[width=#1]{Figures/#4.pdf}
	\hspace*{\fill}
	\caption{#5}
	\label{#2}
\end{figure*}
}%

\usepackage{verbatim}
\usepackage{amsfonts}
\usepackage{amssymb}
\usepackage{amsmath}
\usepackage{hyperref}
\usepackage{graphicx}
\usepackage{color}

\usepackage{listings}
\begin{document}

\maketitle
\thispagestyle{empty}
\pagestyle{empty}

\begin{abstract}
	This paper proposes a novel method for testing observability in Gaussian models using discrete density approximations (deterministic samples) of (multivariate) Gaussians. Our notion of observability is defined by the existence of the maximum a posteriori estimator. In the first step of the proposed algorithm, the discrete density approximations are used to generate a single representative design observation vector to test for observability. In the second step, a number of carefully chosen design observation vectors are used to obtain information on the properties of the estimator. By using measures like the variance and the so-called local variance, we do not only obtain a binary answer to the question of observability but also provide a quantitative measure.
\end{abstract}\

\section{Introduction}

Models that express Gaussian random variables as a function of desired parameters $\mathbf{\Omega}$ and a Gaussian disturbance term are interesting to consider.
In the following, we denote the $d$-dimensional standard normal Gaussian distribution by $\mathcal{N}_d(\mathbf{0},I_d)$. The $d$-dimensional Gaussian distribution with mean $\bmu$ and covariance matrix $\Sigma$ is denoted by $\mathcal{N}_d(\bmu,\Sigma)$.

One of the simplest models is
\begin{equation*}
	Z_t=\Omega_1+\sqrt{\Omega_2} \cdot \epsilon_t, t=1,\ldots,T
\end{equation*}
for $\Omega_1,\Omega_2\in \R$, $\epsilon_t \sim \mathcal{N}(0,1), t=1,\ldots,T$, i.i.d., for estimating mean $\Omega_1$ and variance $\Omega_2$ of a population. In general, the equation for $Z_t$ can be extended to the form $Z_t = g(\mathbf{\Omega}, \epsilon_t) \text{ for } t=1,\ldots,T$.
Of course, it is of interest to know whether this type of model is observable, i.e., if the parameters $\mathbf{\Omega}$ can be estimated from a given observation vector $\mathbf{Z}_T=(Z_1,\ldots,Z_T)^{\top}$. Furthermore, we want to provide a quantitative measure of observability.

We introduce a novel numerical method that investigates observability by using discrete approximations of the density of $\mathcal{N}_d(\mathbf{0},I_d)$ defined in \cite{hanebeckLocalizedCumulativeDistributions2008a} and \cite{steinbringSmartSamplingKalman2016} to study the behavior of the model when optimal design observation vectors are utilized.
The idea of the discrete density approximations is to approximate the continuous density of $\mathcal{N}_d(\mathbf{0},I_d)$ by a discrete distribution that takes on $M$ point-symmetric values with equal probabilities. These values are collected in the set $\mathcal{S}_M^d$ and are called discrete density approximations or deterministic samples. For performing the approximation, we need a distance measure between a continuous distribution and a discrete distribution. This distance measure is defined via the so-called localized cumulative distribution function defined in \cite{hanebeckLocalizedCumulativeDistributions2008a}. Minimizing the distance of $\mathcal{N}_d(\mathbf{0},I_d)$ and the discrete distribution with values in $\mathcal{S}_M^d$ leads to an optimal discrete density approximation.

Another approach would be to simply use random observation vectors. However, that approach is more inefficient and the result depends on the realization. Further, the design observation vectors can lead to exact estimates of the desired parameters with already one realization which is not possible for random observation vectors.

Our approach provides an option to test for observability if analytic calculations are very complex or infeasible. Models where analytic calculations are possible can be used to check our approach.
With our technique it is not only possible to answer the question of observability but also to provide a quantitative measure of observability. This is sometimes called dynamical observability \cite{aguirreStructuralDynamicalSymbolic2018a}.

As we only consider time-independent parameters, the term observability could also be replaced by the term identifiability. However, in the future, the approach described in this paper will also be used for state space models to simultaneously check identifiability of the parameters and observability of the states. Following \cite{hamelinObservabilityIdentifiabilityEpidemiology2021} and seeing parameters as time-independent states, we combine the two and call it observability. Hence, we also use the term observability in this paper.

Identifiability and observability have been considered a lot in the literature already. In \cite{petersIdentifiabilityGaussianStructural2014} for example, penalized maximum likelihood estimation is used for Gaussian models with equal disturbance variances. As the models we consider here are special cases of state space models with constant state, also the literature on the observability of state space models is of interest. In \cite{barfootStateEstimationRobotics2017}, the posterior is maximized over the states. In Chapter 7 of \cite{durbinTimeSeriesAnalysis2012}, maximum likelihood estimation is used to estimate the time-independent parameters. Chapter 13 in \cite{durbinTimeSeriesAnalysis2012} introduces a Bayesian approach to obtain estimates for the parameters by calculating the posterior mean. \cite{walterIdentifiabilityStateSpace1982} gives an extensive overview of the identifiability of state space models.

The remainder of the paper is organized as follows. In \autoref{Sec:Definitions}, we define the considered models and our notion of observability. In \autoref{Sec:Toy} it is shown how observability can be tested when analytic calculations are possible. Then we propose a numeric approach in \autoref{Alterantive_No_Latent} when analytic calculations are complex or infeasible. We end with some examples in \autoref{Sec:Examples} and conclude in \autoref{Sec:Conclusion}.

\section{Definitions} \label{Sec:Definitions}

\subsection{Definition of the Model} \label{Sec:DefModel}

The general model we consider is of the form
\begin{align*}
	Z_t &= g(\mathbf{\Omega}, \epsilon_t) \text{ for } t=1,\ldots,T,
\end{align*}
where $\mathbf{\Omega}$ is a set of time-independent parameters. $Z_t$ is a scalar random observation at time $t$ and $g$ is a function describing how the observation is determined from the parameters $\mathbf{\Omega}$ and disturbances $(\epsilon_t)_{t=1,\ldots,T}$. We assume that $(\epsilon_t)_{t=1,\ldots,T}$ are i.i.d. standard normal disturbances.

Given a realization $\mathbf{z}_T=(z_1,\ldots,z_T)^{\top}$ of the random observation vector $\mathbf{Z}_T=(Z_1,\ldots,Z_T)^{\top}$ from time step $1$ to $T$, we want to estimate the parameters $\mathbf{\Omega}$. If an estimator for $\mathbf{\Omega}$ exists, we denote it by $\hat{\mathbf{\Omega}}(\mathbf{Z}_T)$ and the estimates by $\hat{\mathbf{\Omega}}(\mathbf{z}_T)$.

Furthermore, we define the posterior $\boldsymbol\ell(\mathbf{\Omega}|\mathbf{z}_T)$. In the examples we consider, we use non-informative priors on the parameters, which means that likelihood and posterior are equivalent. However, if prior information is available, it can be incorporated.

We also require the log-posterior denoted by
\begin{equation*}
	L(\mathbf{\Omega}|\mathbf{z}_T)=\log(\boldsymbol\ell(\mathbf{\Omega}\vert \mathbf{z}_T)).
\end{equation*}

\subsection{Our Definition of Observability}\label{Sec:DefObs}

A model is called observable if the parameters can be estimated from a given observation vector.
Citing \cite{balakrishnanIdentificationAutomaticControl1969}, observability ``\textit{can almost always be considered as a problem of finding extrema of functionals. The form of the functional, the extremum of which is to be found, is given by the criterion accepted for the system identification and by the mathematical model of the system}\,''.

Our definition of observability is defined by the existence of the a posteriori estimator and its properties. We always consider the observability for a fixed underlying value $\mathbf{\Omega}^*$.

Given a fixed but arbitrary observation vector $\mathbf{z}_T$ of the model $Z_t=g(\mathbf{\Omega^*},\epsilon_t)$ for finite $T$, we want to find the maximum a posteriori estimate
\begin{equation}\label{Eq:Argmax}
	\hat{\mathbf{\Omega}}(\mathbf{z}_T) = \displaystyle \argmax_{\mathbf{\Omega}} \boldsymbol\ell(\mathbf{\Omega}|\mathbf{z}_T),
\end{equation}
i.e., we are interested in recovering an estimate $\hat{\mathbf{\Omega}}(\mathbf{z}_T)$ of the time-independent parameters $\mathbf{\Omega}^*$.

Given $\mathbf{z}_T$, the ideal case is to have one single maximum of the corresponding posterior function $\boldsymbol\ell(\cdot|\mathbf{z}_T)$. However, in general, there are two additional cases to be considered.

\paragraph*{\textbf{Finitely Many Maxima}}
In the case of two or more distinct maxima of $\boldsymbol\ell(\cdot|\mathbf{z}_T)$, there is no unique maximum a posteriori estimate $\hat{\mathbf{\Omega}}(\mathbf{z}_T)$ of the parameter $\mathbf{\Omega}$ given $\mathbf{z}_T$, but it might be possible to find a unique maximum of the posterior when constraining the parameter space.
In this paper, we consider local observability, i.e., given an observation vector $\mathbf{z}_T$, we just consider a neighborhood around the true values $\mathbf{\Omega}^*$.
Hence, if necessary, the domain of $\argmax$ in \autoref{Eq:Argmax} is constrained to a neighborhood around the true values $\mathbf{\Omega}^*$. 

\cite{anguelovaObservabilityIdentifabilityNonlinear2007}, \cite{bartosiewiczLocalObservabilityNonlinear1995}, and \cite{hermannNonlinearControllabilityObservability1977} deal with global and local observability but in a deterministic setup. For a linear model, global and local observability are equivalent, and it is not possible to have finitely many maxima. Then, only the following case of having infinitely many maxima is interesting.

\paragraph*{\textbf{Infinitely Many Maxima}} In the case where there is a ridge or a plateau, the posterior does not have a distinct maximum. It is not possible to find the true underlying parameters, but possibly values for functions of parameters.
As an example, consider
\begin{equation}\label{Eq:InfEx}
	Z_t=\Omega_1 + \Omega_2 + \epsilon_t, \epsilon_t \sim \mathcal{N}(0,1) \text{ i.i.d.,}
\end{equation}
where the associated log-posterior with uniform priors is of the form
\begin{equation*}
	L(\Omega_1,\Omega_2|\mathbf{Z}_T)\,\propto-\frac{1}{2}\sum_{t=1}^T(Z_t-(\Omega_1+\Omega_2))^2.
\end{equation*}
Here, no distinct maximum of the posterior can be found. The only information we can obtain is $\{(\Omega_1,\Omega_2): \Omega_1 + \Omega_2 = const.\}$, which means that we obtain information about $\Omega_1 + \Omega_2$ but not about $\Omega_1$ and $\Omega_2$ individually. This can also be seen in the surface plot of the posterior given in \autoref{Ridge}.

\SingleFigureHereSpalte{8cm}{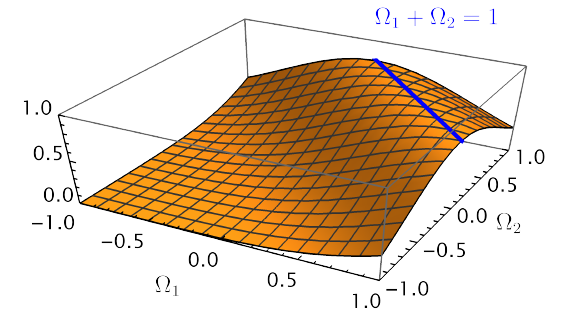}{The posterior of the model given in \autoref{Eq:InfEx}. It is not possible to obtain values of the individual parameters $\Omega_1$ and $\Omega_2$. The values used here are $\Omega_1^*=0.6, \Omega_2^*=0.4$, and $T=2, z_1=1.5, z_2=0.5$.}{Ridge}

In the case of infinitely many maxima for a realization $\mathbf{z}_T$ of $\mathbf{Z}_T$, no distinct maximum is found in \autoref{Eq:Argmax}. If this is the case for all realizations $\mathbf{z}_T$ of $\mathbf{Z}_T$, we call our model unobservable for the underlying value $\mathbf{\Omega}^*$.

Otherwise, we call our model observable for $\mathbf{\Omega}^*$ and obtain an estimator $\hat{\mathbf{\Omega}}(\mathbf{Z}_T)$ corresponding to the estimates $\hat{\mathbf{\Omega}}(\mathbf{z}_T)$ obtained through \autoref{Eq:Argmax}. If a restriction of the parameter space is necessary for at least one realization $\mathbf{z}_T$, we call our model locally observable.

For some models, the estimator $\hat{\mathbf{\Omega}}(\mathbf{z}_T)$ might not be defined for all realizations $\mathbf{z}_T$. An example is
\begin{equation*}
	Z_t=\frac{1}{\Omega}+ \epsilon_t, \epsilon_t \sim \mathcal{N}(0,1) \text{ i.i.d.}
\end{equation*}
The maximum a posteriori estimator when using uninformative priors is
\begin{equation*}
	\hat{\Omega}(\mathbf{Z}_T)=\frac{T}{\sum_{t=1}^T Z_t},
\end{equation*}
which is not defined for a realization $\mathbf{z}_T$ with $\sum_{t=1}^Tz_t=0$.

The estimator should be consistent, i.e., $\hat{\mathbf{\Omega}}(\mathbf{Z}_T) \stackrel{P}{\rightarrow} \mathbf{\Omega}^*$ for $T \rightarrow \infty$, where $\mathbf{\Omega}^*$ is the true underlying value.

For an observable model, we also want to know how well it is observable.
The first way to do this is to consider the variance of the estimator, i.e., $\Var\!\big(\hat{\mathbf{\Omega}}(\mathbf{Z}_T)\big)$.
Another way is to consider the so-called local variance of the parameter estimates $\hat{\mathbf{\Omega}}_j(\mathbf{z}_T), j=1,\ldots,|\Omega|$. We call this $\LVar\!\big(\hat{\mathbf{\Omega}}_j(\mathbf{z}_T)\big)$ for a fixed $\mathbf{z}_T$. Compared to the variance of the estimator where all possible observation vectors are considered, we consider here how distinct the maximum of the log-posterior is, given one observation vector $\mathbf{z}_T$. The local variance is defined in detail in the next paragraph.
Furthermore, it is interesting how many time steps $T$ are required to find a local maximum, i.e., to obtain an estimate, and if a maximum can be found for all $\mathbf{z}_T$.

\subsection{Local Variance}\label{Sec:LocVar}

Given an observation vector $\mathbf{z}_T$, we want to approximate the posterior $\boldsymbol\ell(\cdot|\mathbf{z}_T)$ by a Gaussian density in $\mathbf{\Omega}$ at $\hat{\mathbf{\Omega}}(\mathbf{z}_T)$ if it exists. We use the second-order Taylor polynomial $\tilde{L}$ of the log-posterior $L(\cdot|\mathbf{z}_T)$ around $\hat{\mathbf{\Omega}}(\mathbf{z}_T)$. Then, $\tilde{L}$ can be interpreted as the logarithm of a Gaussian density of $\mathcal{N}_{|\mathbf{\Omega}|}(\bmu^{\mathbf{\Omega}},\Sigma^{\mathbf{\Omega}})$, i.e.,
\begin{equation*}
	\tilde{L}(\mathbf{\Omega})=-\frac{1}{2}\left(\mathbf{\Omega}-\bmu^{\mathbf{\Omega}}\right)^{\top}\cdot \left(\Sigma^{\mathbf{\Omega}}\right)^{-1} \cdot \left(\mathbf{\Omega}-\bmu^{\mathbf{\Omega}}\right)+C
\end{equation*}
for a constant C.
If $\hat{\mathbf{\Omega}}(\mathbf{z}_T)$ is a maximum of $L(\cdot|\mathbf{z}_T)$, we have
\begin{align*}
	\tilde{L}(\mathbf{\Omega})=&-\frac{1}{2}\big(\mathbf{\Omega}-\hat{\mathbf{\Omega}}(\mathbf{z}_T)\big)^{\top}\\
	&\cdot \big(-\mathbf{H}_{L}\big(\hat{\mathbf{\Omega}}\left(\mathbf{z}_T\right)\big)\big) \cdot \big(\mathbf{\Omega}-\hat{\mathbf{\Omega}}(\mathbf{z}_T)\big)+C,
\end{align*}
i.e., $\bmu^{\mathbf{\Omega}}=\hat{\mathbf{\Omega}}(\mathbf{z}_T)$ and $\Sigma^{\mathbf{\Omega}}=-\mathbf{H}_{L}^{-1}\big(\hat{\mathbf{\Omega}}(\mathbf{z}_T)\big)$, where $\mathbf{H}_{L}(\cdot)$ is the Hessian of $L(\cdot|\mathbf{z}_T)$ with respect to $\mathbf{\Omega}$.

For ease of notation, $\mathbf{H}_{L}^{-1}\big(\hat{\mathbf{\Omega}}(\mathbf{z}_T)\big)=\mathbf{H}_{L}^{-1}\big(\hat{\mathbf{\Omega}}(\mathbf{z}_T)|\mathbf{z}_T\big)$, $L(\hat{\mathbf{\Omega}}(\mathbf{z}_T))=L(\hat{\mathbf{\Omega}}(\mathbf{z}_T)|\mathbf{z}_T)$, and the same for the derivatives.

For
\begin{equation*}
	Y_j \sim \mathcal{N}_{1}\Big(\hat{\mathbf{\Omega}}_j\left(\mathbf{Z}_T\right),-\mathbf{H}_{L}^{-1}\big(\hat{\mathbf{\Omega}}\left(\mathbf{Z}_T\right)\big)_{jj}\Big)
\end{equation*}
and given $\mathbf{z}_T$, the local variances of the parameter estimates $\hat{\mathbf{\Omega}}_j\left(\mathbf{z}_T\right)$ are then defined by the diagonal entries
\begin{align}\label{Eq:DefLVar}
	\LVar&\big(\hat{\mathbf{\Omega}}_j\left(\mathbf{z}_T\right)\big):=\Var\!\left(Y_j|\mathbf{Z}_T=\mathbf{z}_T\right)\nonumber\\
	&=\Sigma^{\mathbf{\Omega}}_{jj}=-\mathbf{H}_{L}^{-1}\!\big(\hat{\mathbf{\Omega}}\left(\mathbf{z}_T\right)\!\big)_{jj}, \,j=1,\ldots,|\mathbf{\Omega}|.
\end{align}

\autoref{LocVar} illustrates the local variance in the one-dimensional case $\Omega \in \R$ for a fixed realization $\mathbf{z}_T$. In the left plot, the log-posterior $L(\cdot|\mathbf{z}_T)$ is plotted in blue together with the Taylor approximation $\tilde{L}$ in yellow. The right plot shows the two functions when taking the exponential respectively. Hence, we see the posterior in blue and $\exp(\tilde{L})$ in yellow. The variance and mean of the density given by $\tilde{\boldsymbol\ell} \propto \exp(\tilde{L})$ are of interest to us.

\DoubleFigureHere{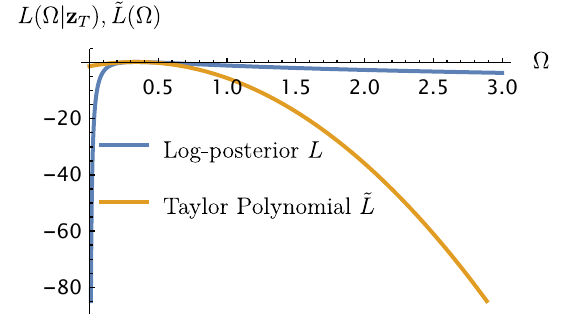}{Illustration of the Taylor polynomial for fixed $\mathbf{z}_T$. In the left plot, the log-posterior $L(\cdot|\mathbf{z}_T)$ is plotted in blue together with the Taylor polynomial $\tilde{L}$ in yellow. The right side shows the two functions when taking the exponential. Hence, we see the posterior in blue and $\exp(\tilde{L})$ in yellow. The variance and mean of the density given by $\tilde{\boldsymbol\ell} \propto \exp(\tilde{L})$ are of interest to us.}

We now show the calculations for a one-dimensional parameter.

\paragraph*{\textbf{One Dimension}} For a one-dimensional parameter $\Omega$ with log-posterior $L(\cdot|\mathbf{z}_T)$, the second-order Taylor polynomial of $L(\cdot|\mathbf{z}_T)$ around $\hat{\Omega}\left(\mathbf{z}_T\right)$ is of the form
\begin{align*}
	\tilde{L}(\Omega)&=L\big(\hat{\Omega}(\mathbf{z}_T)\big)+L'\big(\hat{\Omega}\left(\mathbf{z}_T\right)\!\big)\big(\Omega-\hat{\Omega}\left(\mathbf{z}_T\right)\!\big)\\
	&+\frac{1}{2}L''\big(\hat{\Omega}\left(\mathbf{z}_T\right)\big)\big(\Omega-\hat{\Omega}\left(\mathbf{z}_T\right)\big)^2\\
	&=-\frac{1}{2}\frac{\big(\Omega-\hat{\Omega}\left(\mathbf{z}_T\right)\big)^2}{-\frac{1}{L''\left(\hat{\Omega}\left(\mathbf{z}_T\right)\right)}}+C,
\end{align*}
where we assume that $\hat{\Omega}\left(\mathbf{z}_T\right)$ is the maximum a posteriori estimator of $\Omega$ and hence, $L'\big(\hat{\Omega}\left(\mathbf{z}_T\right)\big)=0$. $C$ is a constant. Then the local variance is the variance corresponding to the Gaussian density $\tilde{\boldsymbol\ell} \propto \exp(\tilde{L}(\Omega))$, i.e.,
\begin{equation*}
	\LVar(\hat{\Omega}\left(\mathbf{z}_T\right))=-\frac{1}{L''\left(\hat{\Omega}\left(\mathbf{z}_T\right)\!\right)}.
\end{equation*}

\section{Toy Example} \label{Sec:Toy}
We now consider the example
\begin{equation}\label{Eq:Example}
	Z_t=\sqrt{b} \cdot \epsilon_t, t=1,\ldots,T \text{ with } \epsilon_t \sim \mathcal{N}(0,1) \text{ i.i.d.}
\end{equation}
and show that the estimator can be determined analytically.

Given a random observation vector $\mathbf{Z}_T=(Z_1,\ldots,Z_T)$, we want to estimate $b$. We consider the posterior with uninformative prior for $b$
\begin{equation*}
    \boldsymbol\ell(b|\mathbf{Z}_T) \, \propto \prod_{t=1}^T\frac{1}{\sqrt{b}}\exp\left[-\frac{1}{2}\frac{Z_t^2}{b}\right]
\end{equation*}
and the corresponding log-posterior
\begin{equation*}
    L(b|\mathbf{Z}_T) \, \propto -\frac{1}{2}\frac{1}{b}\sum_{t=1}^TZ_t^2-\frac{T}{2}\log(b).
\end{equation*}

Calculating the maximum a posteriori estimator leads to
\begin{equation*}
    \hat{b}(\mathbf{Z}_T)=\frac{1}{T}\sum_{t=1}^TZ_t^2.
\end{equation*}
Mean and variance of this estimator can easily be determined by
\begin{equation*}
    \E[\hat{b}(\mathbf{Z}_T)]=b 
\end{equation*}
and
\begin{equation}\label{Eq:Varb}
    \Var\!\big(\hat{b}(\mathbf{Z}_T)\big)=\frac{2}{T}b^2.
\end{equation}

To obtain the local variance, we first have to determine the second-order Taylor polynomial $\tilde{L}$ of the log-posterior $L(\cdot|\mathbf{Z}_T)$ around $\hat{b}(\mathbf{Z}_T)$, which yields
\begin{align*}
    \tilde{L}(b)&= L\big(\hat{b}(\mathbf{Z}_T)\big)+\frac{1}{2}L''\big(\hat{b}(\mathbf{Z}_T)\big)\cdot\big(b-\hat{b}(\mathbf{Z}_T)\big)^2\\
    &=-\frac{1}{2}\frac{T^3}{2\left(\sum_{t=1}^TZ_t^2\right)^2}\left(b-\frac{\sum_{t=1}^TZ_t^2}{T}\right)^2+C.
\end{align*}
The first derivative is zero because we evaluate it at the maximum $\hat{b}(\mathbf{Z}_T)$.
For given $\mathbf{z}_T$, this leads to the local variance
\begin{equation}\label{Eq:LVarb}
    \LVar\!\big(\hat{b}(\mathbf{z}_T)\big)=\frac{2\left(\sum_{t=1}^TZ_t^2\right)^2}{T^3}=\frac{2}{T}\hat{b}^2(\mathbf{z}_T).
\end{equation}
\autoref{Fig_Algorithm1} shows the general algorithm when analytical calculations are possible.

\begin{algorithm}[hh]
    \DontPrintSemicolon
    \smallskip
    \KwIn{\begin{itemize}
            \item Posterior $\boldsymbol\ell$
            \item Number of time steps $T$
         \end{itemize}}
    \smallskip

	Determine $\hat{\mathbf{\Omega}}(\mathbf{Z}_T)$ analytically by maximizing $\boldsymbol\ell(\cdot|\mathbf{Z}_T)$
		 \smallskip

    \uIf{$\hat{\mathbf{\Omega}}(\mathbf{Z}_T)$ is a maximum of $\boldsymbol\ell(\cdot|\mathbf{Z}_T)$}{
        Determine variance and mean of $\hat{\mathbf{\Omega}}(\mathbf{Z}_T)$
		}
	\Else{
		Not observable!
	}
    \caption{Testing Observability Analytically}
    \label{Fig_Algorithm1}
\end{algorithm}

\section{Our Proposed Method} \label{Alterantive_No_Latent}

In \autoref{Sec:Toy}, calculating the estimator is done analytically.
As this is not always possible or associated with long calculations, we now introduce a numerical approach using a similar idea as introduced in \autoref{Fig_Algorithm1}.

\subsection{Key Idea} \label{Sec:KeyIdea}

As described in Section \ref{Sec:DefObs}, the problem is how to investigate if a model has a maximum a posteriori estimator. Furthermore, we are interested in the properties of the estimator.

The observability and the properties of the estimator are always considered for fixed parameters $\mathbf{\Omega}^*$. Then, increasing numbers of time steps $T$ are considered.

The algorithm is separated into two parts. In Part I, we consider one representative observation vector to test for observability. In Part II, we determine mean and variance of the estimator. In both parts, we do not consider arbitrary observation vectors, but we carefully choose so-called design observation vectors.
This means that the behavior of the model can be worse for real-life data.

The design observation vectors are generated using the approach defined in \cite{hanebeckLocalizedCumulativeDistributions2008a} and \cite{steinbringSmartSamplingKalman2016}. In general, we approximate the density of $\mathcal{N}_d(\mathbf{0},I_d)$ using a discrete distribution with $M$ point masses. These point masses are then chosen as design values for the disturbances $\epsilon_1,\ldots,\epsilon_T$ of the considered model to generate design values for the observed data (design observation vectors). In the following, we explain the procedure for both parts in detail.

\paragraph*{\textbf{Part I}} First, we generate one representative design observation vector $\tilde{\mathbf{z}}_T$ to check for observability. Using $\tilde{\mathbf{z}}_T$, we maximize the log-posterior $L(\cdot|\tilde{\mathbf{z}}_T)$ numerically and check if a maximum $\hat{\mathbf{\Omega}}(\tilde{\mathbf{z}}_T)$ can be found (see Section \ref{Sec:Checks}). If so, the model is observable and an estimator $\hat{\mathbf{\Omega}}(\mathbf{Z}_T)$ exists.
If not,
we always consider Part II as well to check for several design observation vectors. If no maximum is found, we conclude that the model is unobservable. This takes care of models where the estimator exists but is not defined in $\tilde{\mathbf{z}}_T$.

To determine the representative design observation vector, we want only one design disturbance vector $(\tilde{\epsilon}_1,\ldots,\tilde{\epsilon}_T)$ for $(\epsilon_1,\ldots,\epsilon_T)$. For $T=2$, we select $(\tilde{\epsilon}_1,\tilde{\epsilon}_2)=(-1,1)$ with mean zero and variance one. This corresponds to approximating the density of $\mathcal{N}(0,1)$ with $M=2$ point masses. The approach can be applied to $T>2$ by using the fact that $\epsilon_t \sim \mathcal{N}(0,1), t=1,\ldots,T$, i.i.d. Then, we approximate the density of $\mathcal{N}(0,1)$ by the discrete distribution with $M=T$ point masses chosen by the Algorithm in \cite{hanebeckLocalizedCumulativeDistributions2008a} and \cite{steinbringSmartSamplingKalman2016}. 

The associated design observation vector is defined by
\begin{align*}
	\tilde{z}_t &= g(\mathbf{\Omega}, \tilde{\epsilon}_t), t=1,\ldots,T.
\end{align*}

\paragraph*{\textbf{Part II}} Second, we are interested in the expected value and variance of the estimator $\hat{\mathbf{\Omega}}(\mathbf{Z}_T)$. For this, we generate $K$ design observation vectors $\tilde{\mathbf{z}}_T^k, k=1,\ldots,K$, and determine the empirical mean and variance from the resulting estimates
\begin{equation*}
	\hat{\mathbf{\Omega}}(\tilde{\mathbf{z}}_T^k)=\argmax_{\mathbf{\Omega}} \boldsymbol\ell(\mathbf{\Omega}|\tilde{\mathbf{z}}_T^k).
\end{equation*}
For every maximum, we run through a series of checks defined in Section \ref{Sec:Checks}. There may be observable models for which we do not find an estimate for some design observations as explained in Section \ref{Sec:DefModel}. However, we only consider observation vectors that can occur with a high probability, so we might not find all observation vectors $\mathbf{z}_T$ for which no estimate exists in theory.

In order to obtain $K$ design disturbance vectors, we use
$(\epsilon_1,\ldots,\epsilon_T)^{\top} \sim \mathcal{N}_T(\mathbf{0},I_T).$
Then, we approximate the density of $\mathcal{N}_T(\mathbf{0},I_T)$ by a density with a discrete distribution with $M=K$ point mass vectors.
We obtain $K$ different design disturbance vectors $(\tilde{\epsilon}_1^k,\ldots,\tilde{\epsilon}_T^k), k=1,\ldots,K$, for $(\epsilon_1,\ldots,\epsilon_T)$.

The associated design observation vectors are defined by
\begin{align*}
	\tilde{z}_t^k &= g(\mathbf{\Omega}, \tilde{\epsilon}_t^k), t=1,\ldots,T, k=1,\ldots,K.
\end{align*}

\subsection{Check Maxima}\label{Sec:Checks}

The numerical maximization is performed using L-BFGS optimization over all unknown parameters $\mathbf{\Omega}$. After obtaining the result from the optimization procedure, we have to do some checks in order to be sure that we have found a maximum.

The optimization procedure searches for minima, hence we use the function $-2L$. Then, given an observation vector $\mathbf{z}_T$, we check the following properties of the result $\hat{\mathbf{\Omega}}(\mathbf{z}_T)$. The thresholds are chosen based on simulation results.
\paragraph*{\textbf{Gradients}} Are the gradients of $-2L\big(\hat{\mathbf{\Omega}}(\mathbf{z}_T)\big)$ small enough? For observable models, the gradients are in the magnitude of $10^{-9}$. The threshold we use is $10^{-5}$.
\paragraph*{\textbf{Hessian}} Is the Hessian $H_{-2L}\big(\hat{\mathbf{\Omega}}(\mathbf{z}_T)\big)$ positive definite? If yes, we have a minimum of $-2L(\cdot|\mathbf{z}_T)$ and hence a maximum of the posterior $\boldsymbol\ell$.
\paragraph*{\textbf{Eigenvalues}} Is the result not a ridge? The ratio of the smallest eigenvalue to the largest eigenvalue of $H_{-2L}(\hat{\mathbf{\Omega}}(\mathbf{z}_T))$ should not be too small. We want a ratio of $> 0.00001$, in observable examples it is in the magnitude of $0.0001$.
\paragraph*{\textbf{Local Variance}} Is the result not a plateau? A plateau would mean that the local variances are infinity. Hence, we always have to check that this is not the case.

\autoref{Fig_Algorithm2} summarizes the proposed algorithm.

\begin{algorithm}[hh]
    \DontPrintSemicolon
    \smallskip
    \KwIn{\begin{itemize}
            \item Posterior $\boldsymbol\ell$
            \item Number of time steps $T$
         \end{itemize}}
    \smallskip

	\textbf{Part I} Use repr. design observation vector $\tilde{\mathbf{z}}_T=(\tilde{z}_1,\ldots,\tilde{z}_T)$

	\Indp Get maximum $\hat{\mathbf{\Omega}}(\tilde{\mathbf{z}}_T)$ of $\boldsymbol\ell(\cdot|\tilde{\mathbf{z}}_T)$ numerically

    \uIf{$\hat{\mathbf{\Omega}}(\tilde{\mathbf{z}}_T)$ fulfills the checks in \ref{Sec:Checks}}{
        Model observable, Go to Part II
		}
	\Else{
		Go to Part II to check if not observable
	}

	\smallskip

	\Indm \textbf{Part II} Use $K$ design observation vectors $\tilde{\mathbf{z}}_T^k=(\tilde{z}_1^k,\ldots,\tilde{z}_T^k)$

	\Indp Get up to $K$ estimates $\hat{\mathbf{\Omega}}(\tilde{\mathbf{z}}_T^k)$ numerically

	Get emp. variance and mean of $\hat{\mathbf{\Omega}}(\mathbf{Z}_T)$ using $\hat{\mathbf{\Omega}}(\tilde{\mathbf{z}}_T^k)$

    \caption{Testing Observability Numerically}
    \label{Fig_Algorithm2}
\end{algorithm}

\subsection{Example for  Design Observation Vectors}
\SingleFigureHereSpalte{8cm}{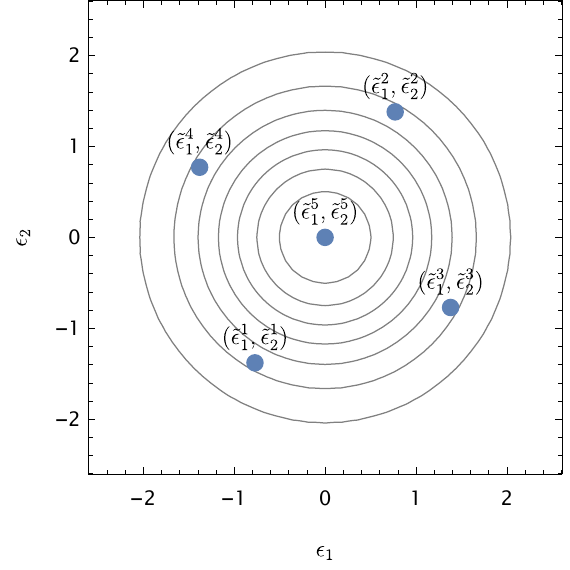}{The $M=K=5$ design disturbance vectors together with the contour plots of the standard normal density.}{DetSamplingLCD25}
We illustrate the idea for the model in \autoref{Eq:Example} and $T=2$ time steps. We want to investigate the observability when design values of $(Z_1=\sqrt{b}\epsilon_1, Z_2=\sqrt{b}\epsilon_2)$ are given.

First, we want to obtain the representative design observation vector. For that, we use the fact that $\epsilon_1, \epsilon_2 \sim \mathcal{N}(0,1)$ i.i.d., so we approximate the density of $\mathcal{N}(0,1)$ with $M=T=2$ point masses and get the design disturbance vectors $\left(\tilde{\epsilon}_1=1, \tilde{\epsilon}_2=-1\right)$. The design observation vector is then given by
\begin{equation*}
	(\tilde{z}_1,\tilde{z}_2)=(\sqrt{b}\tilde{\epsilon}_1,\sqrt{b}\tilde{\epsilon}_2)=(\sqrt{b},-\sqrt{b}).
\end{equation*}

Now, we determine the maximum a posteriori estimate numerically by maximizing $L(\cdot|(\tilde{z}_1,\tilde{z}_2))$. The result is the same as the one determined from the analytic estimator 
\begin{equation*}
	\frac{1}{T}\sum_{t=1}^T\tilde{z}_t^2=\frac{1}{2}(b+b)=b.
\end{equation*}
The existence confirms observability.

Now, we want to know the behavior of the estimator using $K$ different design observation vectors. In this example, we choose $K=5$ but in order to determine the empirical mean and variance, we later choose $K=2000$. To obtain the $K=5$ design disturbance vectors, we use
\begin{equation*}
	(\epsilon_1, \epsilon_2) \sim \mathcal{N}_2(\mathbf{0},I_2).
\end{equation*}
Then, we approximate the density of $\mathcal{N}_2(\mathbf{0},I_2)$ by $M=K=5$ point mass vectors $(\tilde{\epsilon}_1^k,\tilde{\epsilon}_2^k)_{k=1,\ldots,5}$. They are illustrated in \autoref{DetSamplingLCD25}.

Using these design disturbance vectors, we obtain the design observation vectors
\begin{equation*}
	(\tilde{z}_1^k,\tilde{z}_2^k)=(\sqrt{b}\tilde{\epsilon}_1^k,\sqrt{b}\tilde{\epsilon}_2^k), k=1,\ldots,5.
\end{equation*}
They are used to obtain $5$ estimates of $b$ by maximizing $L(\cdot|(\tilde{z}_1^k,\tilde{z}_2^k))$ for $k=1,\ldots,5$ individually.

\iftrue
\bgroup
\def\arraystretch{1.5}
\begin{table*}[t]
\renewcommand\thetable{II}
\centering
\begin{tabular}{|c|c|c|c|c|c|}
	\hline
	& Emp. $\E\!\left[\hat{b}(\mathbf{Z}_T)\right]$ & $\Var\left(\hat{b}(\mathbf{Z}_T)\right)$ & Emp. $\Var\!\left(\hat{b}(\mathbf{Z}_T)\right)$ & Mean of $\LVar\!\left(\hat{b}(\tilde{\mathbf{z}}_T^k)\right)$ & Mean of Num. $\LVar\!\left(\hat{b}(\tilde{\mathbf{z}}_T^k)\right)$ \\
	\hline
	T &$\bar{\mu}=\frac{\sum_{k=1}^K\hat{b}(\tilde{\mathbf{z}}_T^k)}{K}$ & $\frac{2}{T}b^2$ & $\frac{\sum_{k=1}^K\left(\hat{b}(\tilde{\mathbf{z}}_T^k)-\bar{\mu}\right)^2}{K-1}$ & $\frac{\sum_{k=1}^K\frac{2}{T}(\hat{b}(\tilde{\mathbf{z}}_T^k))^2}{K}$ & $\frac{\sum_{k=1}^K -\frac{1}{L''\left(\hat{b}(\tilde{\mathbf{z}}_T^k)\right)}}{K}$ \\
	\hline
	4 & 0.8 & 0.320 & 0.31 & 0.475 & 0.475 \\
	\hline
	12 & 0.8 & 0.107 & 0.099 & 0.123 & 0.123 \\
	\hline
	20 & 0.8 & 0.064 & 0.0576 & 0.0698 & 0.0698 \\
	\hline
\end{tabular}
\caption{\label{Table:Ka1} The empirical mean and variance of $\hat{b}(\mathbf{Z}_T)$. They are determined using the estimates $\hat{b}(\tilde{\mathbf{z}}_T^k)$ for example \ref{Example:b} with the $2000$ design observation vectors $\left(\tilde{\mathbf{z}}_T^k\right)$. Further, the mean of the local variances of $\hat{b}(\tilde{\mathbf{z}}_T^k)$ is given. The variance can be determined by \autoref{Eq:Varb} or by the empirical variance. The local variances can be determined by \autoref{Eq:LVarb} or by using \autoref{Eq:DefLVar}}.
\end{table*}
\egroup
\fi

\iftrue
\ThreeFigureDifferentHere{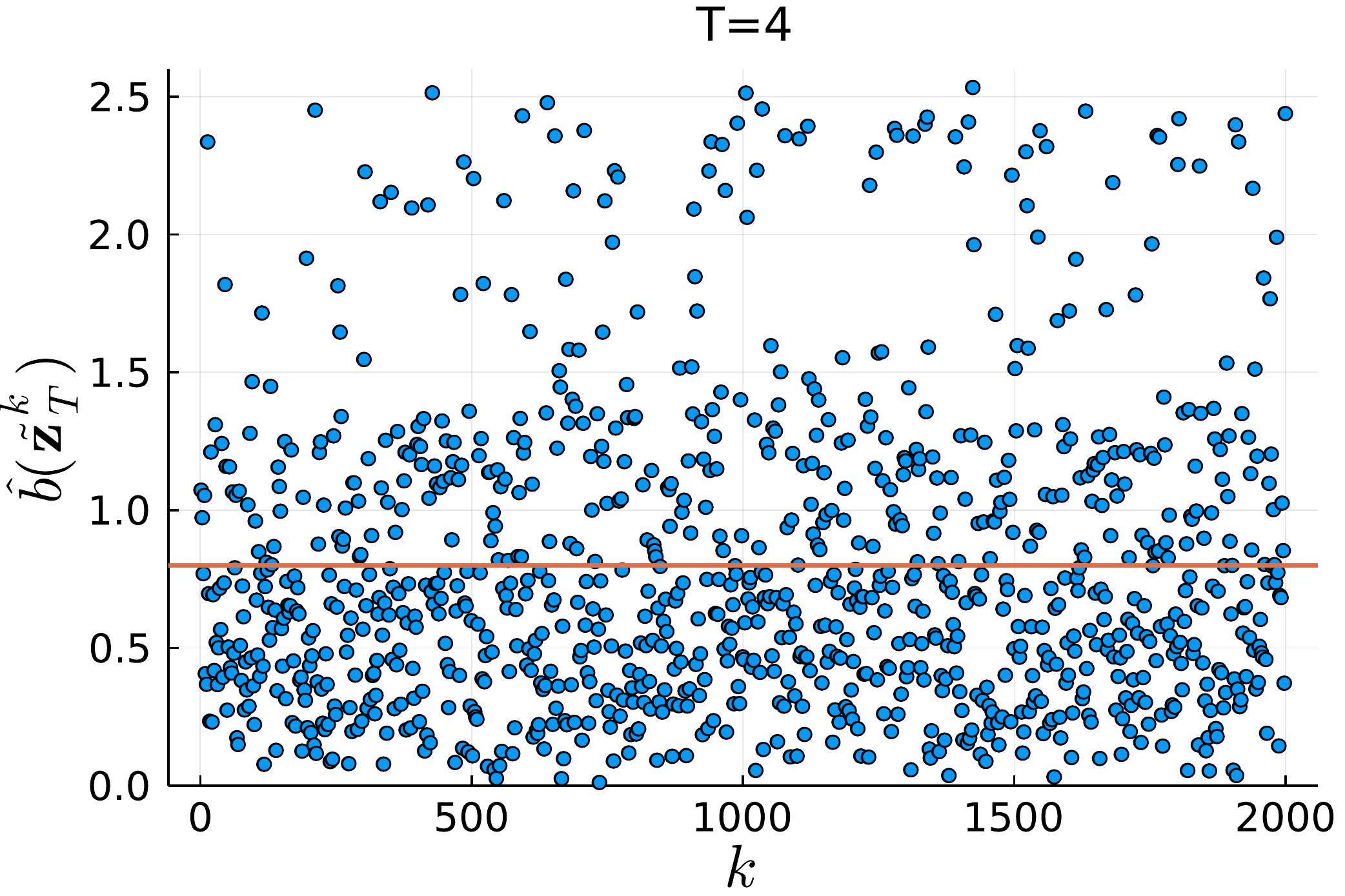}{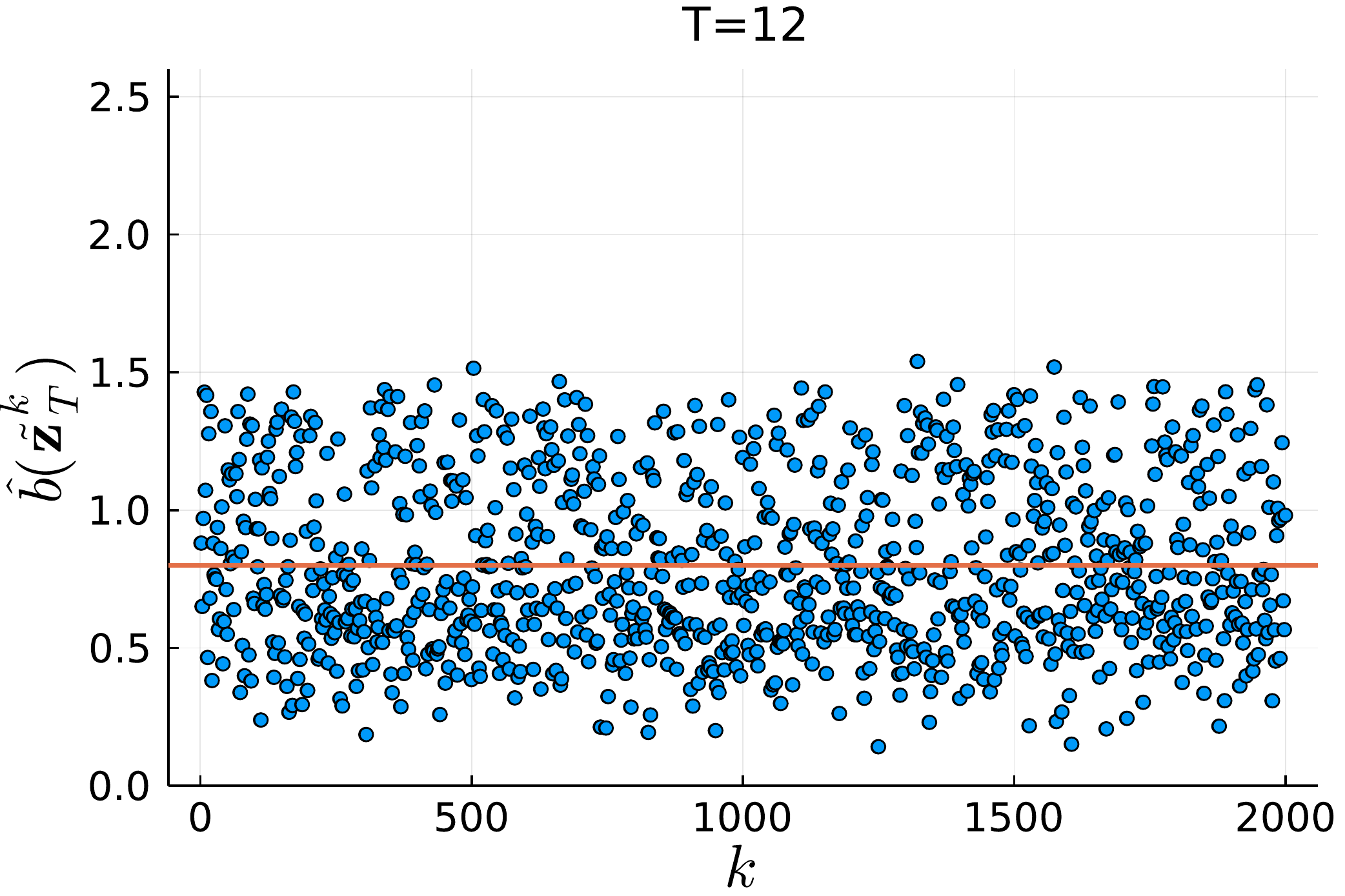}{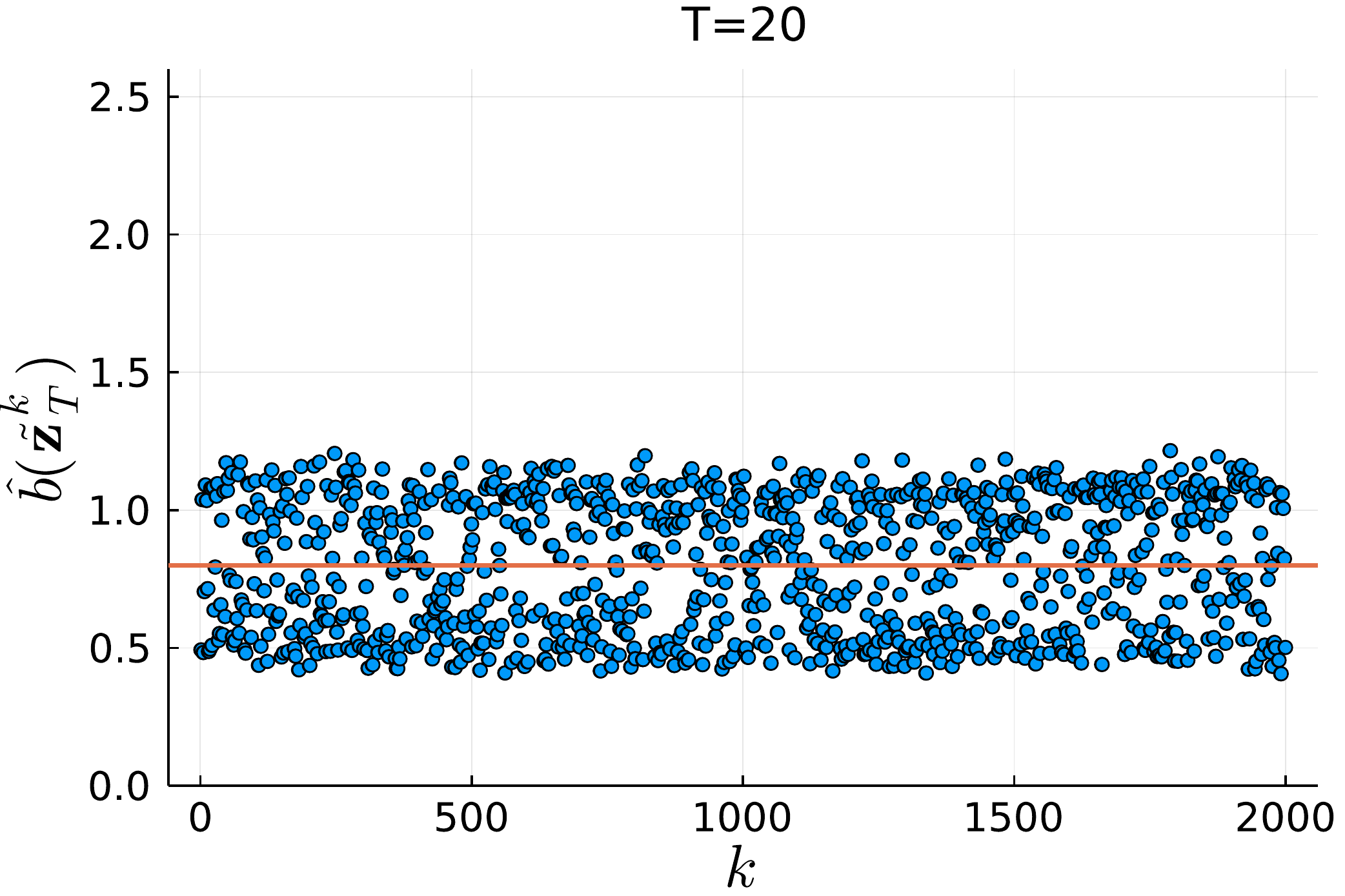}{The estimates $\hat{b}(\tilde{\mathbf{z}}_T^k), k=1,\ldots,2000$, for different choices of $T$. On the $y$-axis, the estimate $\hat{b}(\tilde{\mathbf{z}}_T^k)$ that is obtained by maximizing $L(\cdot|\tilde{\mathbf{z}}_T^k)$ for the respective iteration is plotted. The orange line shows $b^*=0.8$.}
\fi

\bgroup
\def\arraystretch{1.5}
\begin{table}[t]
	\renewcommand\thetable{I}
\centering
\begin{tabular}{|c|c|c|c|c|c|}
	\hline
	 &  & $\LVar\!\left(\hat{b}(\tilde{\mathbf{z}}_T)\right)$ & Num. $\LVar\!\left(\hat{b}(\tilde{\mathbf{z}}_T)\right)$\\ 
	\hline
	$T$ & $\hat{b}(\tilde{\mathbf{z}}_T)$ & $\frac{2}{T}\hat{b}(\tilde{\mathbf{z}}_T)^2$ & $-\frac{1}{L''\left(\hat{b}(\tilde{\mathbf{z}}_T)\right)}$\\
	\hline
	4 & 0.8 & 0.320 & 0.320\\
	\hline
	12 & 0.8 & 0.107 & 0.107\\
	\hline
	20 & 0.8 & 0.064 & 0.064\\
	\hline
\end{tabular}
\caption{\label{Table:Repa} The values and the local variance of the estimates $\hat{b}(\tilde{\mathbf{z}}_T)$ for example \ref{Example:b} using the representative design observation vector $\tilde{\mathbf{z}}_T$. The local variance of $\hat{b}(\tilde{\mathbf{z}}_T)$ can either be determined by \autoref{Eq:LVarb} or by using \autoref{Eq:DefLVar}.}
\end{table}
\egroup

\section{Examples} \label{Sec:Examples}

\subsection{Model with Unknown Variance $b$} \label{Example:b}

Consider again the model from \autoref{Eq:Example}.
In general, we are interested in the parameter $b$, given an observation vector $\textbf{z}_T$. We consider this example by using the algorithm introduced in \autoref{Alterantive_No_Latent}. For this, we set the true underlying parameter of $b$ to $b^*=0.8$, i.e., we examine the observability for this value of $b$. Further, we consider the number of time steps $T=4, T=12$, and $T=20$. We do not have to increase this number further because we get very satisfactory results for even these small numbers of time steps.

The log-posterior is of the form
\begin{equation*}
	L(b|\textbf{z}_T)\, \propto-\frac{1}{2}\frac{1}{b}\sum_{t=1}^Tz_t^2-\frac{T}{2}\log(b).
\end{equation*}
As we use an uninformative prior for $b$, the log-posterior is the same as the log-likelihood.

For Part I of the algorithm, we determine the representative design observation vector $\tilde{\mathbf{z}}_T$ for the three choices of $T$ and determine the corresponding estimates $\hat{b}(\tilde{\mathbf{z}}_T)$ by maximizing $\boldsymbol\ell(\cdot|\tilde{\mathbf{z}}_T)$ (see \autoref{Table:Repa}). Further, the local variances of these estimates are determined. The three maxima fulfill the requirements defined in Section \ref{Sec:Checks}. As can be expected from the representative design observation vector, the estimate is always precise. The local variance of $\hat{b}(\tilde{\mathbf{z}}_T)$ is decreasing with increasing $T$, i.e., the uncertainty of the maximum decreases.

For Part II, we now consider $K=2000$ different design observation vectors $(\tilde{z}_1^k,\ldots,\tilde{z}_T^k), k=1,\ldots,K$, to study the properties of the estimator. In \autoref{Table:Ka1}, the empirical mean and variance of $\hat{b}(\mathbf{Z}_T)$ are shown. They are determined using the estimates $\hat{b}(\tilde{\mathbf{z}}_T^k)$ given the $K$ design observation vectors $\tilde{\mathbf{z}}_T^k, k=1,\ldots,K$. Further, the mean of the local variances of $\hat{b}(\tilde{\mathbf{z}}_T^k)$ is given. All $3 K$ maxima pass the checks defined in Section \ref{Sec:Checks}.
There is a deviation of the empirical variance to the true variance $\Var\!\big(\hat{b}(\mathbf{Z}_T)\big)$. This deviation decreases by increasing $K$: For $T=20$ for example, we obtain an estimate of $0.0607$ for $K=5000$ and $0.0628$ for $K=10000$.

It can be seen that we have an unbiased consistent estimator. The variance of the estimator $\hat{b}(\mathbf{Z}_T)$ and the local variances of the individual estimates $\hat{b}(\tilde{\mathbf{z}}_T^k)$ decrease as expected and hence the quality of observability increases.

In \autoref{b_0.8_4}, the estimates $\hat{b}(\tilde{\mathbf{z}}_T^k)$ are plotted versus the iteration number $k$ for $k=1,\ldots,2000$.

\iftrue
\bgroup
\def\arraystretch{1.5}
\begin{table*}[t]
\renewcommand\thetable{IV}
	\centering
	\begin{tabular}{|c|c|c|c|c|}
		\hline
		& Emp. $\E\!\left[\hat{\mathbf{\Omega}}(\mathbf{Z}_T)\right]$ & Emp. $\Var\!\left(\hat{\mathbf{\Omega}}(\mathbf{Z}_T)\right)$ & Mean of $\LVar\!\left(\hat{\mathbf{\Omega}}(\tilde{\mathbf{z}}_T^k)\right)$ \\
		\hline
		T& $(\bar{\mu}^1,\bar{\mu}^2)=\frac{\sum_{k=1}^K(\hat{a}(\tilde{\mathbf{z}}_T^k),\hat{b}(\tilde{\mathbf{z}}_T^k))}{K}$ & $\frac{\sum_{k=1}^K\left((\hat{a}(\tilde{\mathbf{z}}_T^k)-\bar{\mu}^1)^2,(\hat{b}(\tilde{\mathbf{z}}_T^k)-\bar{\mu}^2)^2\right)}{K-1}$  & $\frac{\sum_{k=1}^K \left(\left(-\mathbf{H}_{L}^{-1}(\hat{\mathbf{\Omega}}(\tilde{\mathbf{z}}_T^k)\right)_{11},\left(-\mathbf{H}_{L}^{-1}(\hat{\mathbf{\Omega}}(\tilde{\mathbf{z}}_T^k)\right)_{22}\right)}{K}$ \\
		\hline
		4 & (0.6, 0.3) & (0.1, 0.0584) & (0.075, 0.0742)\\
		\hline
		12 & (0.6, 0.367) & (0.0334, 0.0228) & (0.0306, 0.0262)\\
		\hline
		20 & (0.6, 0.38) & (0.02, 0.0138) &(0.019, 0.0158) \\
		\hline
	\end{tabular}
\caption{\label{Table:Kab1} The empirical mean and variance of $\hat{\mathbf{\Omega}}(\mathbf{Z}_T)$. They are determined using the estimates $(\hat{a}(\tilde{\mathbf{z}}_T^k),\hat{b}(\tilde{\mathbf{z}}_T^k))$ for model \ref{Example:ab} with the $2000$ design observation vectors $\tilde{\mathbf{z}}_T^k, k=1,\ldots,2000$. Further, the mean of the local variances of $(\hat{a}(\tilde{\mathbf{z}}_T^k),\hat{b}(\tilde{\mathbf{z}}_T^k))$ is given. The local variances are determined using \autoref{Eq:DefLVar}.}
\end{table*}
\egroup

\ThreeFigureDifferentHere{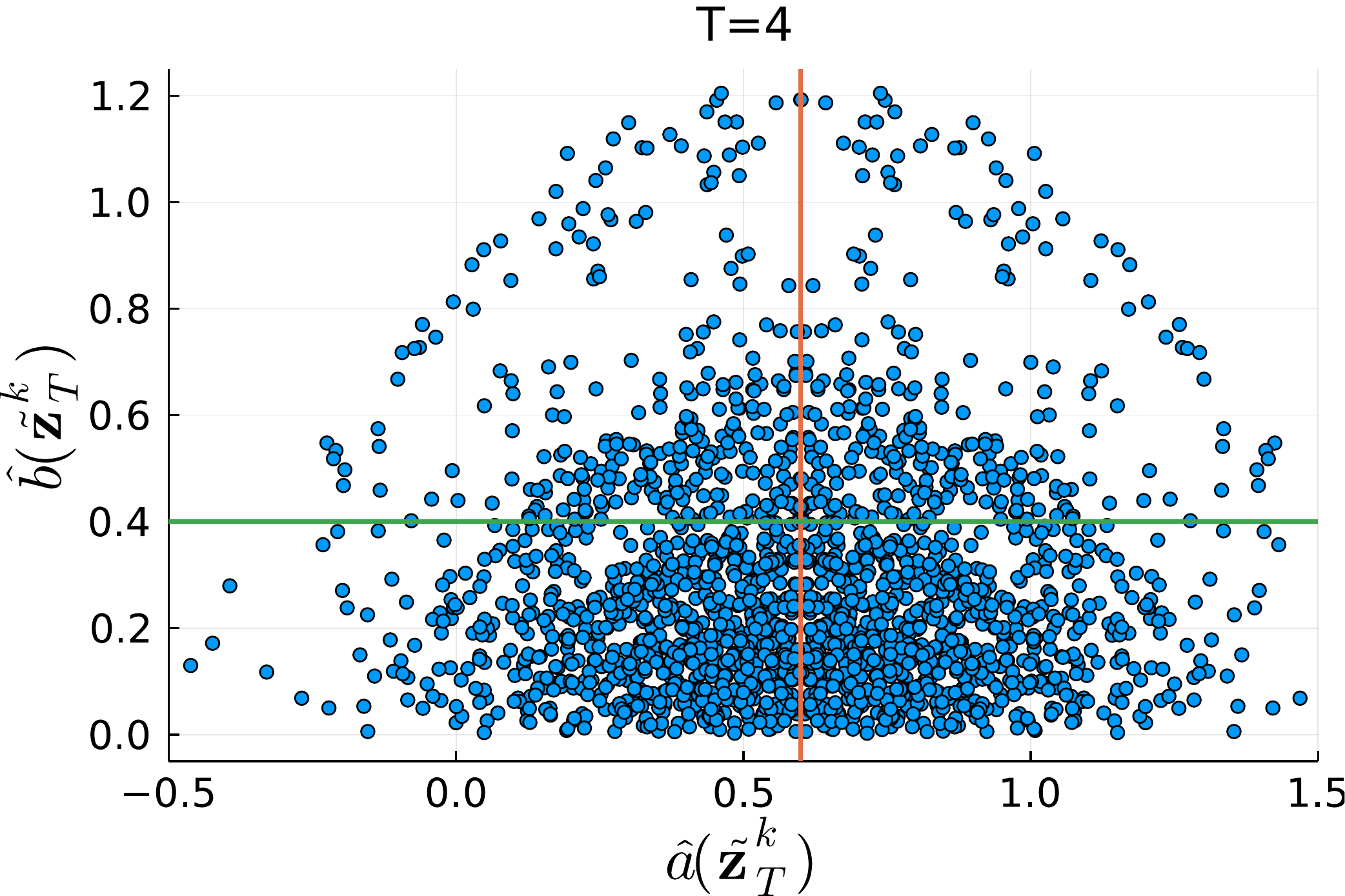}{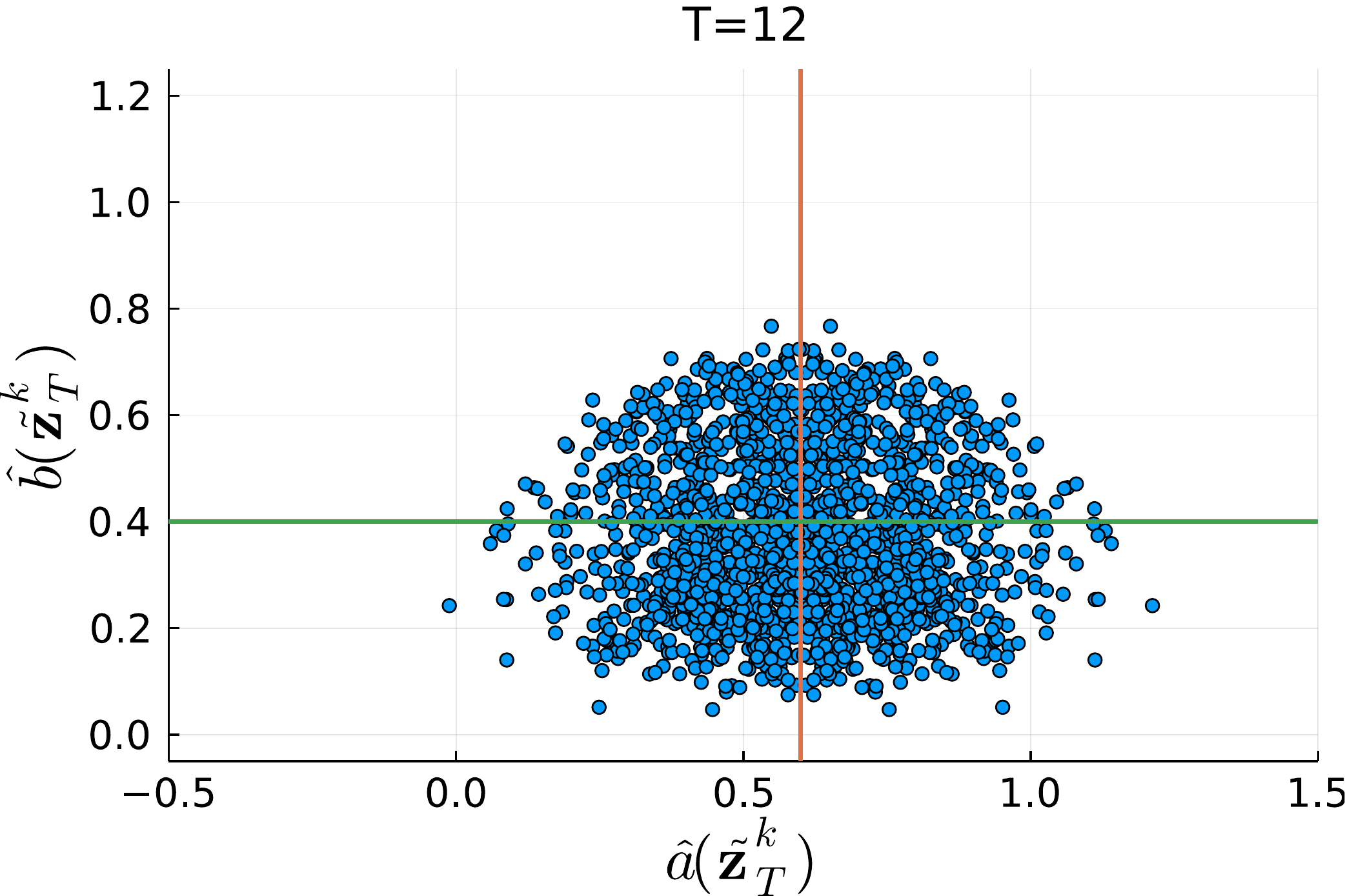}{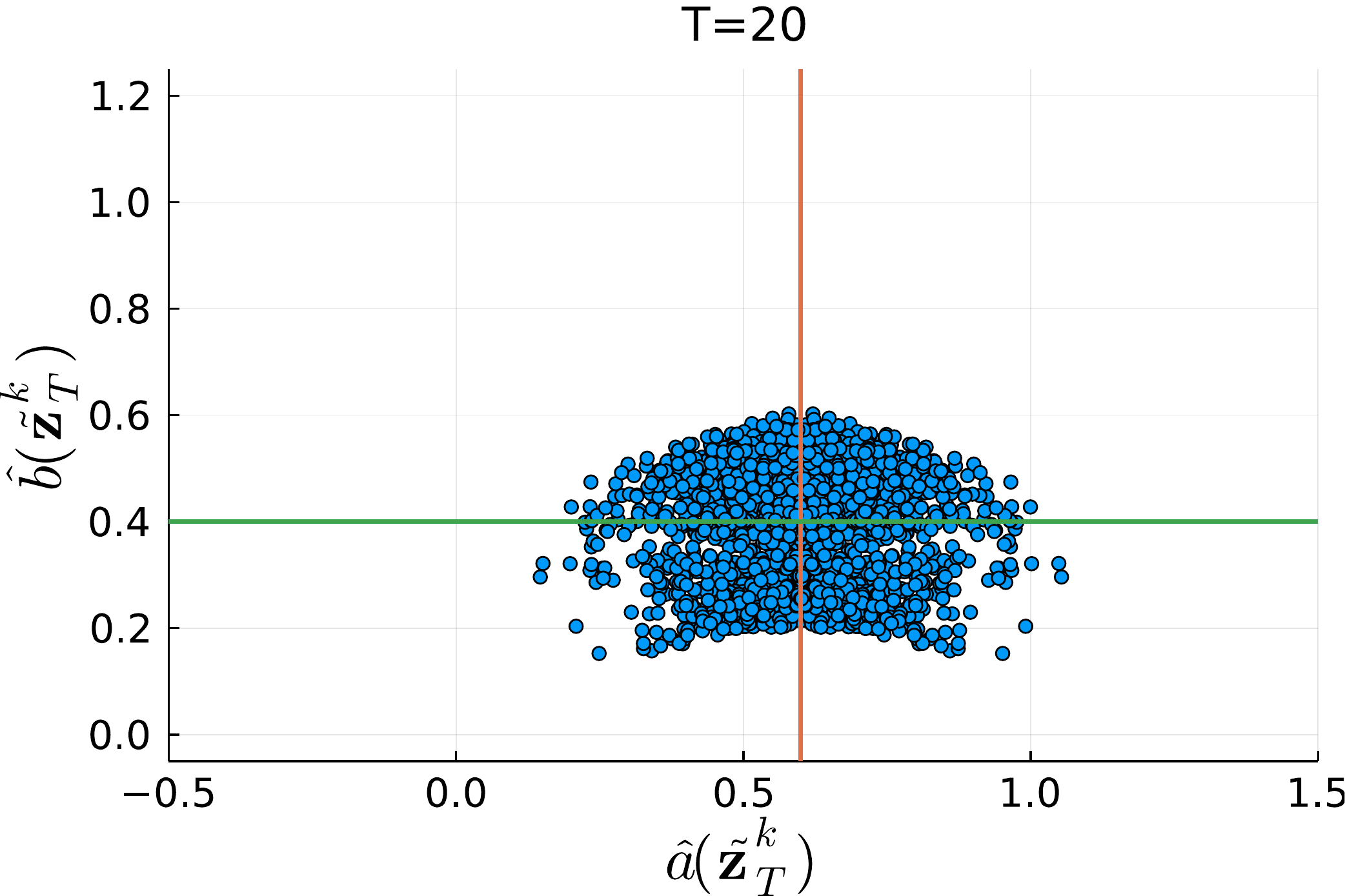}{The estimates $(\hat{a}(\tilde{\mathbf{z}}_T^k),\hat{b}(\tilde{\mathbf{z}}_T^k)), k=1,\ldots,2000$, for different choices of $T$. The estimates are obtained by maximizing $L(\cdot| \tilde{\mathbf{z}}_T^k)$ individually for $k=1,\ldots,2000$. The orange line shows $a^*=0.6$ while the green line shows $b^*=0.4$.}
\fi

\bgroup
\def\arraystretch{1.5}
\begin{table}[t]
	\renewcommand\thetable{III}
\centering
\begin{tabular}{|c|c|c|c|}
	\hline
	 &  & $\left(\LVar\!\left(\hat{a}(\tilde{\mathbf{z}}_T)\right),\LVar\!\left(\hat{b}(\tilde{\mathbf{z}}_T)\right)\right)$\\
	\hline
	$T$ & $\left(\hat{a}(\tilde{\mathbf{z}}_T),\hat{b}(\tilde{\mathbf{z}}_T)\right)$ & $\text{diag}\left((-\mathbf{H}_{L}^{-1}\!\left(\hat{\mathbf{\Omega}}(\tilde{\mathbf{z}}_T)\right)\!\right)$\\
	\hline
	4 & (0.6, 0.4) & (0.1, 0.08)\\
	\hline
	12  & (0.6, 0.4) & (0.0333, 0.0267)\\
	\hline
	20 & (0.6, 0.4)& (0.02, 0.016)\\
	\hline
\end{tabular}
\caption{\label{Table:Repab} The values and the local variance of the estimates $(\hat{a}(\tilde{\mathbf{z}}_T),\hat{b}(\tilde{\mathbf{z}}_T))$ for model \ref{Example:ab} using the representative design observation vector $\tilde{\mathbf{z}}_T$. The local variance is determined using \autoref{Eq:DefLVar}.}
\end{table}
\egroup

Using a smaller value for $b^*$, the variances of the estimator and the local variances are smaller than for $b^*=0.8$. Hence, the quality of observability is higher for smaller true values of $b$. The results are available upon request.


\subsection{Model with Unknown Mean $a$ and Unknown Variance $b$} \label{Example:ab}

Consider
\begin{equation*}
Z_t=a+\sqrt{b} \cdot \epsilon_t, t=1,\ldots,T,  \text{ with } \epsilon_t \sim \mathcal{N}(0,1) \text{ i.i.d.}
\end{equation*}
Here, $\mathbf{\Omega}=(a,b)$ is unknown and of interest to us for a given observation vector $\textbf{z}_T$. We set the true underlying parameters to $(a^*,b^*)=(0.6,0.4)$ and consider $T \in \{4,12,20\}$ again.

The log-posterior is of the form
\begin{equation*}
	L(a,b|\textbf{z}_T)\, \propto-\frac{1}{2}\frac{1}{b}\sum_{t=1}^T(z_t-a)^2-\frac{T}{2}\log(b),
\end{equation*}
where we use uninformative priors for $a$ and $b$ again.

Here, it is important to note that we have a bias on the estimator. It can be shown analytically that the theoretical expected value is
\begin{equation*}
	\E\big[(\hat{a}(\mathbf{Z}_T),\hat{b}(\mathbf{Z}_T))\big]=\left(a^*,\frac{T-1}{T}b^*\right).
\end{equation*}
This bias is also visible in the numerical outcome Emp. $\E\!\big[\hat{\mathbf{\Omega}}(\mathbf{Z}_T)\big]$ given in \autoref{Table:Kab1}.

In Part I, we determine the representative design observation vector $\tilde{\mathbf{z}}_T$, $T\in \{4,12,20\}$ and the corresponding maxima $(\hat{a}(\tilde{\mathbf{z}}_T),\hat{b}(\tilde{\mathbf{z}}_T))$ of $\boldsymbol\ell(\cdot|\tilde{\mathbf{z}}_T)$, and study their local variances (see \autoref{Table:Repab}). The requirements in Section \ref{Sec:Checks} are fulfilled.

For Part II, we consider $K=2000$ design observation vectors. In \autoref{Table:Kab1}, the empirical mean and variance of $\hat{\mathbf{\Omega}}(\mathbf{Z}_T)$ are given. They are determined using the estimates $(\hat{a}(\tilde{\mathbf{z}}_T^k),\hat{b}(\tilde{\mathbf{z}}_T^k))$ given the $K$ design observation vectors $\tilde{\mathbf{z}}_T^k, k=1,\ldots,K$. Further, the mean of the local variances of $(\hat{a}(\tilde{\mathbf{z}}_T^k),\hat{b}(\tilde{\mathbf{z}}_T^k)), k=1,\ldots,K$, is given. All $3K$ maxima fulfill the requirements in Section \ref{Sec:Checks}. As in the example before, the variance of the estimator and the local variances decrease. \autoref{a_b_0.6_0.4_4} shows the estimates $(\hat{a}(\tilde{\mathbf{z}}_T^k),\hat{b}(\tilde{\mathbf{z}}_T^k)), k=1,\ldots,K$, for $a$ and $b$.

\subsection{Some more Examples}
We always consider $\epsilon_t \sim \mathcal{N}(0,1) \text{ i.i.d.}$ Then, for the models 
\begin{equation*}
Z_t=\frac{a}{b}+\sqrt{a} \cdot \epsilon_t, t=1,\ldots,T
\end{equation*}
and 
\begin{equation*}
	Z_t=\frac{a}{b}+\sqrt{a\cdot b} \cdot \epsilon_t, t=1,\ldots,T
\end{equation*}
it can be shown that they are observable. The results are available upon request.
On the other hand,
\begin{equation}\label{Eq:Model1}
	Z_t=\frac{a}{b}+\sqrt{\frac{a}{b}} \cdot \epsilon_t, t=1,\ldots,T 
\end{equation}
and
\begin{equation}\label{Eq:Model2}
	Z_t=a\cdot b+\epsilon_t, t=1,\ldots,T
\end{equation}
are not observable because the posterior has a ridge. The two posterior functions are illustrated in \autoref{Ridgeadurchb} and \autoref{Ridgeamalb}.

\SingleFigureHereSpalte{8cm}{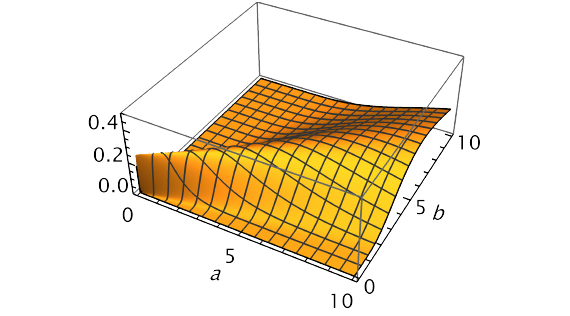}{The posterior of the model defined in \autoref{Eq:Model1}. The values used here are $a^*=0.6, b^*=0.4$, and $\tilde{\epsilon}_1=1, \tilde{\epsilon}_2=-1$.}{Ridgeadurchb}

\SingleFigureHereSpalte{8cm}{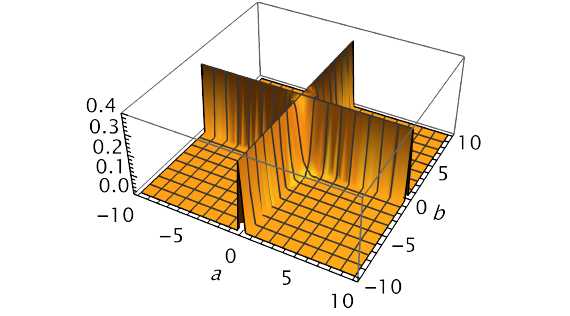}{The posterior of the model defined in \autoref{Eq:Model2}. The values used here are $a^*=0.6, b^*=0.4$, and $\tilde{\epsilon}_1=1, \tilde{\epsilon}_2=-1$.}{Ridgeamalb}

\section{Conclusions and Outlook} \label{Sec:Conclusion}

We introduced a novel numerical method to investigate the observability of Gaussian models. This offers an alternative to sometimes complex or even infeasible analytic calculations. Using measures such as the variance and the local variance of the estimator, we do not only answer the question of observability but also obtain a quantitative measure of the quality of observability.

The next step is to consider state space models. The goal is to show observability of copula state space models defined in \cite{kreuzerBayesianNonlinearState2022} and \cite{kreuzerBayesianMultivariateNonlinear2019}.

\balance
\bibliographystyle{abbrv}
\bibliography{References}
\end{document}